\documentstyle{mn}
\ifnfsstwo
  
  \newcommand{\rmn}[1] {{\mathrm #1}}

\fi

\ifnfssone
  \newmathalphabet{\mathit}
    \addtoversion{normal}{\mathit}{cmr}{m}{it}
    \addtoversion{bold}{\mathit}{cmr}{bx}{it}
    
    \newcommand{\rmn}[1] {{\mathrm #1}}

\fi

\ifoldfss
  
  \newcommand{\rmn}[1] {{\rm #1}}

\fi

\loadboldmathitalic
\loadboldgreek

\onecolumn

\input epsf.sty

\newcommand{\bmi}{\bmath}

\newcommand{\ddlnr}{\partial / \partial \ln r_{12} |_{\triangle}}
\newcommand{\sdelta}{\delta^s}
\newcommand{\sdeltaLG}{\delta^{s {\rmn {LG}}}}
\newcommand{\sdeltaast}{\delta^{s \ast}}
\newcommand{\rdelta}{\delta}
\newcommand{\rdeltaast}{\delta^\ast}
\newcommand{\sxi}{\xi^s}
\newcommand{\sxiLG}{\xi^{s {\rmn {LG}}}}
\newcommand{\rxi}{\xi}
\newcommand{\sXi}{\Xi^s}
\newcommand{\sN}{N^s}
\newcommand{\rN}{N}
\newcommand{\sP}{P^s}

\newcommand{\sPhi}{\Phi^s}
\newcommand{\rPhi}{\Phi}
\newcommand{\seta}{\eta^s}
\newcommand{\half}{1/2}
\newcommand{\oneandahalf}{3/2}
\newcommand{\ff}{\beta}
\newcommand{\bnabla}{\bf\nabla}
\newcommand{\rxibar}{\bar\rxi}
\newcommand{\rxibarbar}{\skew4\bar{\bar\xi}}
\newcommand{\tildesXi}{\widetilde\sXi}

\title{Spherical redshift distortions}
\author[A. J. S. Hamilton and M. Culhane]
       {A.~J.~S.~Hamilton
        and M.~Culhane \\
        Joint Institute for Laboratory Astrophysics
        and Department of Astrophysical, Planetary and Atmospheric Sciences,\\
        Box 440, University of Colorado, Boulder, CO 80309, USA;
        email: ajsh@dark.colorado.edu, culhane@jila.colorado.edu}
\date{Accepted 1995 July 26.
      Received 1995 July 10;
      in original form 1995 March 14}
\pubyear{1995}
 
\begin{document}
\maketitle
 
\begin{abstract}
Peculiar velocities induce apparent line of sight displacements of
galaxies in redshift space, distorting the pattern of clustering.
On large scales,
the amplitude of the distortion
yields a measure of the dimensionless linear growth rate
$\ff \approx \Omega^{0.6} /b$
of fluctuations,
where $\Omega$ if the cosmological density and $b$ is the bias factor.
To make the maximum statistical use of the data in a wide-angle redshift survey,
and for the greatest accuracy,
the spherical character of the distortion needs to be treated properly,
rather than in the simpler plane-parallel approximation.
In the linear regime,
the redshift space correlation function
is described by a spherical distortion operator acting on
the true correlation function.
It is pointed out here that there exists an operator,
which is essentially the logarithmic derivative with respect to pair separation,
which both commutes with the spherical distortion operator,
and at the same time defines a characteristic scale of separation.
The correlation function can be expanded in eigenfunctions
of this operator,
and these eigenfunctions are eigenfunctions of the distortion operator.
Ratios of the observed amplitudes of the eigenfunctions
yield measures of the linear growth rate $\ff$
in a manner independent of the shape of the correlation function.
More generally,
the logarithmic derivative $\partial / \partial \ln r$
with respect to depth $r$,
along with the square $L^2$
and component $L_z$ of the angular momentum operator,
forms a complete set of commuting operators for the spherical distortion
operator acting on the density.
The eigenfunctions of this complete set of operators
are spherical waves about the observer,
with radial part lying in logarithmic real or Fourier space.

\end{abstract}

\begin{keywords}
large-scale structure of Universe --
cosmology: theory
\end{keywords}
\section{Introduction}

Peculiar velocities induce apparent line of sight displacements of
galaxies in redshift space, distorting the pattern of clustering.
In an influential paper,
Kaiser (1987) pointed out that, in the large-scale linear regime,
and in the plane-parallel approximation,
the distortion takes a particularly simple form in Fourier space.
He showed that a wave of amplitude
$\tilde\rdelta ( {\bmi k} )$ appears amplified
in redshift space by a factor $1 + \ff \mu^2$
(in this paper a superscript $s$ is used to distinguish
the redshift space overdensity $\tilde\sdelta$
from the unredshifted overdensity $\tilde\rdelta$,
and likewise the redshift space correlation function $\sxi$
and power spectrum $\sP$ from their unredshifted
counterparts $\xi$ and $P$):
\begin{equation}
\label{Kaiserd}
 \tilde\sdelta ( {\bmi k} ) =
  ( 1 + \ff \mu^2 ) \tilde\rdelta ( {\bmi k} )
 \ .
\end{equation}
Here $\mu \equiv \hat{\bmi z} . \hat{\bmi k}$
is the cosine of the angle between the wavevector ${\bmi k}$ and the line of
sight ${\bmi z}$,
and $\ff$ is the growth rate of growing modes in linear theory,
the dimensionless quantity which solves the linearized continuity equation
${\bmi\nabla} . {\bmi v} + \ff \rdelta = 0$
in units where the Hubble constant is one.
In unbiased standard pressureless Friedmann cosmology,
the linear growth rate $\ff$
depends on the cosmological density parameter $\Omega$ as
(e.g.\ Peebles 1980, equation [14.8])
\begin{equation}
\label{f}
 \ff \approx \Omega^{0.6} \ .
\end{equation}
If the galaxy overdensity $\rdelta$
is linearly biased by a factor $b$
relative to the underlying matter density $\rdelta_{\rmn M}$ of the Universe,
$\rdelta = b \rdelta_{\rmn M}$,
but velocities are unbiased,
then the observed value of $\ff$ is modified to
\begin{equation} 
\label{fb}
 \ff \approx {\Omega^{0.6} \over b}
 \ .
\end{equation}
However,
the assumption of linear bias lacks compelling justification,
and equation (\ref{fb}) is really just an acknowledgment that
the measurement of $\Omega$ may be biased if galaxies do not trace mass.
In general, $\ff$ depends on the adopted cosmological model,
and is a function of $\Omega$,
the cosmological constant $\Lambda$,
bias,
and perhaps other quantities.

Kaiser concluded from equation (\ref{Kaiserd})
that the power spectrum of galaxies
(the Fourier transform of the correlation function)
appears in redshift space amplified by a factor
$( 1 + \ff \mu^2 )^2$:
\begin{equation}
\label{Kaiser}
 \sP ( {\bmi k} ) =
  ( 1 + \ff \mu^2 )^2 P ( k ) \ .
\end{equation}

Kaiser's formulae (\ref{Kaiserd}) and (\ref{Kaiser}) assume
the plane-parallel approximation,
where galaxies are taken to be sufficiently far away from the observer
that the displacements induced by peculiar velocities are effectively parallel.
To date, most studies of large-scale redshift space distortions
have assumed the plane-parallel approximation
(Lilje \& Efstathiou 1989;
McGill 1990;
Loveday et al.\ 1992;
Hamilton 1992, 1993a;
Gramann, Cen \& Bahcall 1993;
Bromley 1994;
Fry \& Gazta\~naga 1994;
Fisher et al.\ 1994a; Fisher 1995;
Cole, Fisher \& Weinberg 1994, 1995).
In the plane-parallel approximation one is necessarily restricted
to considering only pairs of galaxies separated by no more than some
angle on the sky.
For example,
Hamilton (1993a) and Cole et al.\ (1995)
restricted the opening angle to $50^\circ$
as a reasonable compromise between statistical uncertainties
and the error resulting from the plane-parallel approximation.
Cole et al.\ (1994, fig.\ 8)
show from simulations that at this opening angle
the plane-parallel approximation causes $\ff$ to be underestimated
by about 5 per cent.

Properly, however,
the redshift displacements of galaxies are radial about the observer,
not plane-parallel.
A correct treatment of radial distortions,
besides being more accurate for pairs of modest angular separation,
would improve statistics by admitting pairs at large angular separations,
which would be particularly helpful for measurements at the largest scales.
In all sky surveys such as the {\em IRAS} redshift surveys, for example,
one could effectively double
(by going fore and aft instead of just fore)
the effective scale over which redshift distortions can be measured.
The need for greater accuracy will increase
as redshift surveys grow larger,
and statistical errors decrease.

The first study of spherical redshift distortions
was by Fisher, Scharf \& Lahav (1994b) on the 1.2~Jy redshift survey.
They expanded the density field in spherical harmonics,
windowing the density in the radial direction with Gaussian windows
at several depths.
Heavens \& Taylor (1995)
improved on Fisher et al.'s procedure
by expanding the radial density field in a complete set of spherical waves.
Both sets of authors assumed a prior form of the (unredshifted) power spectrum
in determining a maximum likelihood value of $\ff$.
Ballinger, Heavens \& Taylor (1995) took Heavens \& Taylor's approach
a step further by allowing the power spectrum to vary in six bins.
One curious aspect of these studies
is that they gave values consistently higher,
$\ff = 0.96^{+ 0.20}_{- 0.18}$ (Fisher et al., 1.2~Jy),
$\ff = 1.1 \pm 0.3$ (Heavens \& Taylor, 1.2~Jy),
and $\ff = 1.04 \pm 0.3$ (Ballinger et al., 1.2~Jy),
than values obtained assuming the plane-parallel approximation,
$\ff = 0.69^{+ 0.28}_{- 0.24}$ (Hamilton 1993a, 2~Jy),
$\ff = 0.45^{+ 0.27}_{- 0.18}$ (Fisher et al.\ 1994a, 1.2~Jy),
and
$\ff = 0.52 \pm 0.15$ and
$\ff = 0.54 \pm 0.3$ (Cole et al.\ 1995, 1.2~Jy and QDOT).
The origin of this discrepancy is not yet clear.

The purpose of the present paper is to set forth a procedure for
measuring $\ff$ from spherical distortions in a manner independent
of the form of the power spectrum.
The basic idea is to represent the correlation function in
eigenfunctions of the spherical distortion operator.

It is instructive to see how this works in the plane-parallel case.
In general, the linear redshift distortion equation (\ref{Kaiserd})
is an operator equation.
In ordinary space, for example, it is
\begin{equation}
 \sdelta ( {\bmi r} ) =
  [ 1 + \ff (\partial / \partial z )^2 \nabla^{-2} ] \, \rdelta ( {\bmi r} )
\end{equation}
where $\nabla^{-2}$ is the inverse Laplacian.
A key advantage of Kaiser's formulation
is that Fourier modes are eigenfunctions of the (plane-parallel)
distortion operator.
This is intimately related to the circumstance that the ${\rmn i} \bnabla$ operator,
whose eigenfunctions define the Fourier modes,
commutes with the distortion operator.

The other key aspect of Kaiser's formula (\ref{Kaiser})
is that the unredshifted power spectrum $P (k)$ is a function only of
the absolute value $k$ of the wavevector.
This implies that ratios of the redshift power spectrum
for waves with the same $k$ but different angles to the line of sight
(different $\mu$)
yield measures of $\ff$ independent of the
amplitude $P (k)$ of the power spectrum.
This argument identifies $k^2$,
or equivalently the operator $- \nabla^2$,
as playing a special role,
which is, roughly stated,
that $k^2$ `defines a scale' of separation.

The first aim, then, of this paper is to identify an operator,
analogous to $k^2$ in the plane-parallel case,
which both commutes with the spherical distortion operator,
and at the same time `defines a scale' of separations.
This is done in Section 3,
the spherical distortion operator having been derived in Section 2.
The procedure is carried a step further in Section 4,
which presents a complete set of commuting operators for the spherical
distortion operator.
The eigenfunctions of this complete set of commuting operators
are logarithmic spherical waves.
Section 5 discusses how to measure $\ff$.
The conclusions are summarized in Section 6.
\section{The Spherical Distortion Operator}
\label{SDO}

This section derives the operator equations (\ref{bdelta}) and (\ref{xi}) which
relate the redshift density, hence correlation function,
to the real density and correlation function
in the linear regime.
We assume the standard gravitational instability picture
in the standard pressureless Friedmann cosmology (e.g.\ Peebles 1980).
The derivation treats correctly the spherical character
of the redshift space distortions about the observer,
so structures may be nearby and may subtend a large angle.
It is nevertheless necessary
to exclude a local region of the Universe about the Milky Way,
both to avoid local bias (Section \ref{bias}),
and to ensure the linear requirement that peculiar velocities
be small compared to the distance to the observed structures,
equation (\ref{v<r}).
The derivation here is similar to that of Kaiser (1987, section 2).

In this subsection,
the frame of reference is taken to be stationary,
that is, the frame of reference of the Cosmic Microwave Background (CMB).
Subsection \ref{LGsec} below discusses
the transformation to the Local Group frame.

Let $s$ denote the observed redshift distance (in the CMB frame) to a galaxy,
and $r$ its true distance from the observer.
The two distances differ by the line of sight peculiar velocity $v$
of the observed galaxy:
\begin{equation}
\label{s}
 s = r + v
\end{equation}
in units where the Hubble constant is unity.
The number $\sN ( {\bmi s} ) {\rmn d}^3 \! s$
of galaxies observed in an interval
${\rmn d}^3 \! s$ of redshift space in a redshift survey is related to the
real space number density $\rN ( {\bmi r} )$ by number conservation:
\begin{equation}
\label{N}
 \sN ( {\bmi s} ) {\rmn d}^3 \! s = \rN ( {\bmi r} ) {\rmn d}^3 \! r \ .
\end{equation}
An unbiased estimate of the true galaxy overdensity
$\rdelta ( {\bmi r} )$ at position ${\bmi r}$ is given by
$1 + \rdelta ( {\bmi r} ) = \rN ( {\bmi r} ) / \rPhi (r)$,
where $\rPhi (r)$ is the selection function,
the expected number density of galaxies at depth $r$
given the selection criteria of the survey.
As emphasized by Fisher et al.\ (1994b),
the selection function $\rPhi (r)$ in a flux-limited survey is
a function of the true distance $r$, not of the redshift distance $s$.
Thus one would be inclined to estimate the overdensity $\sdelta ( {\bmi s} )$
in redshift space by
$1 + \sdelta ( {\bmi s} ) = \sN ( {\bmi s} ) / \rPhi (r)$.
However, this estimate of the overdensity $\sdelta ( {\bmi s} )$
requires knowing not only the true selection function $\rPhi (r)$,
but also the true distance $r$ to the galaxy at redshift distance $s$.
This can be done,
but it requires a full reconstruction of the deredshifted density field,
which is not the philosophy of the present paper.
We assume here instead that one measures,
by some standard technique (Binggeli, Sandage \& Tammann 1988),
a selection function
$\sPhi (s)$ in redshift space,
and that the overdensity $\sdelta ( {\bmi s} )$ in redshift space
is then defined by
$1 + \sdelta ( {\bmi s} ) = \sN ( {\bmi s} ) / \sPhi (s)$.
We distinguish here between the measured redshift space selection function
$\sPhi (s)$ and the true selection function $\rPhi (r)$,
though, as argued below, the two in fact agree to linear order.
Thus the relation between the observed redshift space overdensity
$\sdelta ( {\bmi s} )$
and the true overdensity $\rdelta ( {\bmi r} )$ is
\begin{equation}
\label{Phin}
 \sPhi (s) [ 1 + \sdelta ( {\bmi s} ) ] s^2 ds =
  \rPhi (r) [ 1 + \rdelta ( {\bmi r} ) ] r^2 dr \ .
\end{equation}

With equation (\ref{s}),
equation (\ref{Phin}) rearranges to
\begin{equation}
\label{n}
 1 + \sdelta ( {\bmi s} ) =
  {\rPhi (r) \over \sPhi ( r + v )}
  \left( 1 + {v \over r} \right)^{-2}
  \left( 1 + {\partial v \over \partial r} \right)^{-1}
  [ 1 + \rdelta ( {\bmi r} ) ]
\end{equation}
which, with some small differences,
is Kaiser's (1987) equation (3.2).
In linear theory, the peculiar velocity of the galaxy along the line of sight is
\begin{equation}
\label{v}
 v ( {\bmi r} ) =
  - \ff {\partial \over \partial r} \nabla^{-2} \rdelta ( {\bmi r} )
\end{equation}
where $\nabla^{-2}$ denotes the inverse Laplacian.
Thus to linear order in the overdensity $\rdelta ( {\bmi r} )$,
and with the additional condition that peculiar velocities $v$ be
much smaller than the distance $r$ to the structures being observed,
\begin{equation}
\label{v<r}
 v \ll r
 \ ,
\end{equation}
equation (\ref{n}) reduces to
[note in particular that $\sdelta ( {\bmi s} ) = \sdelta ( {\bmi r} )$
to linear order]
\begin{equation}
\label{bdelta}
 \sdelta ( {\bmi r} ) =
  \left[ 1
   + \ff \left( {\partial^2 \over \partial r^2}
   + {\alpha (r) \partial \over r \partial r} \right) \nabla^{-2}
   \right] \rdelta ( {\bmi r} )
\end{equation}
where $\alpha (r)$ is the logarithmic slope of
$r^2$ times the redshift space selection function $\sPhi (r)$ at depth $r$:
\begin{equation}
\label{alpha}
 \alpha (r) \equiv {\partial \ln r^2 \sPhi (r) \over \partial \ln r} \ .
\end{equation}
Equation (\ref{bdelta})
is an operator equation relating the redshift space overdensity $\sdelta$
to the real space overdensity $\rdelta$.
The quantity in square brackets in equation (\ref{bdelta})
is the spherical distortion operator.
It is straightforward to transform the equation into Fourier space
(Zaroubi \& Hoffman 1995),
but, unlike the plane-parallel case considered by Kaiser (1987),
the spherical distortion operator does not simplify to an eigenvalue
in Fourier space,
but remains an operator.
Tegmark \& Bromley (1995)
obtain an expression for the Green's function (i.e. the inverse) of the
spherical distortion operator in equation (\ref{bdelta}),
for the particular case $\alpha = 2$.

Deriving the linearized equation (\ref{bdelta}) 
from equation (\ref{n})
involves the approximation
$\sPhi (r) = \rPhi (r)$,
which is valid to linear order, as will now be argued.
In a flux-limited survey,
the selection function $\rPhi (r)$ at depth $r$
is the number density of galaxies luminous enough to be seen to depth $r$,
which is the integrated luminosity function above a threshold.
All modern methods of measuring the selection function
(Binggeli et al.\ 1988)
seek to eliminate dependence on density inhomogeneity,
by assuming that the luminosity function is independent of density,
and constructing the luminosity function by counting relative numbers
of bright and faint galaxies in identical volumes.
If bright and faint galaxies at any point share the same peculiar velocity,
as seems probable in the linear regime,
then the measurement of the selection function in redshift space
should remain unbiased by inhomogeneity.
A systematic difference
between the real and redshift space selection functions
will however result from the fact that
on average peculiar velocities will cause galaxies
to tend to `diffuse' away from depths where the density per unit velocity
$r^2 \rPhi (r)$ is larger, towards regions, both shallower and deeper,
where the density is smaller.
This diffusion will perturb the shape of the luminosity function
observed in redshift space from its true shape,
biasing the measurement of $\sPhi (r)$.
The diffusive flux of galaxies is proportional to the variance
$\langle \Delta v^2 \rangle$
of radial peculiar velocities at depth $r$,
which is of second order in perturbations,
and hence vanishes to linear order.
It follows that the redshift and real selection functions agree to linear
order, $\sPhi (r) = \rPhi (r)$, as claimed.
It is to be noted that in general
the difference between $\sPhi (r)$ and $\rPhi (r)$
is confined only to the monopole mode about the observer,
and only to the longest wavelength modes,
since the difference of the functions is slowly varying with depth $r$.
Thus in any case there is little harm in approximating
$\sPhi (r) = \rPhi (r)$
for all but the fundamental modes.

Fisher et al.\ (1994b) make the clever point that
the observed galaxy density weighted by some window $W ( {\bmi s} )$ is
\begin{equation}
\label{WdN}
 W({\bmi s}) \sN ({\bmi s}) {\rmn d}^3 \! s =
 W ({\bmi s}) \rN ({\bmi r}) {\rmn d}^3 \! r \approx
 W ({\bmi r})
 \left( 1 + {v \over r} {\partial \ln W \over \partial \ln r} \right )
 \rN ({\bmi r}) {\rmn d}^3 \! r
\end{equation}
for sufficiently smooth windows $W ( {\bmi s} )$.
Equation (\ref{WdN}),
which is basically an integration by parts,
allows the effect of redshift distortions to be cast in terms of the window
$W ({\bmi s})$, rather than the selection function $\sPhi (s)$
and the overdensity $\sdelta ( {\bmi s} )$.
However,
our aim here is to
decouple the measurement of the linear growth rate $\ff$ from
the shape of the correlation function.
For this purpose it appears essential to work with the overdensity
$\sdelta ( {\bmi s} )$ deconvolved from the shape of the selection function,
which compels us to work with equation (\ref{bdelta})
rather than equation (\ref{WdN}).

The redshift space correlation function follows from equation (\ref{bdelta})
by taking an ensemble average of products of densities at positions
${\bmi r}_1$ and ${\bmi r}_2$
relative to the observer:
\begin{equation}
\label{xi}
 \sxi ( r_{12} , r_1 , r_2 ) =
  \langle \sdelta ( {\bmi r}_1 ) \sdelta ( {\bmi r}_2 ) \rangle =
  \left[ 1
   + \ff \left( {\partial^2 \over \partial r_1^2}
   + {\alpha (r_1) \partial \over r_1 \partial r_1} \right) \nabla_1^{-2}
   \right]
  \left[ 1
   + \ff \left( {\partial^2 \over \partial r_2^2}
   + {\alpha (r_2) \partial \over r_2 \partial r_2} \right) \nabla_2^{-2}
   \right]
  \rxi ( r_{12} )
\end{equation}
where $r_{12} \equiv | {\bmi r}_1 - {\bmi r}_2 |$ is the pair separation.
Equation (\ref{xi})
is the basic equation
which describes the redshift space correlation function
$\sxi (  r_{12} , r_1 , r_2 )$
in the linear regime
as the result of the spherical distortion operator acting
on the true correlation function $\rxi ( r_{12} )$.
It is a consequence of isotropy about the observer
that the redshift correlation function
$\sxi (  r_{12} , r_1 , r_2 )$
is a function only of the lengths of the sides,
not of the orientation,
of the triangle defined by the observer and the pair of galaxies observed.

\subsection{Local bias}
\label{bias}

Equation (\ref{xi}) is the ensemble average redshift correlation function
observed by stationary observers at random positions in the Universe.
But we are not at a random position:
we reside on a galaxy, the Milky Way.

Consider an ensemble of observers who sit on galaxies,
measuring the correlation function.
The probability that an observer on a galaxy at position 0
sees a pair of galaxies at points 1 and 2
is the probability of finding a triple of galaxies at 0, 1, and 2,
divided by the probability of finding the galaxy at 0.
It follows that the apparent correlation function $\xi_{12}( \mbox{apparent} )$
observed from galaxies at 0 is
\begin{equation}
\label{xiapp}
 \xi_{12}( \mbox{apparent} )
 = \langle ( 1 + \delta_0 ) \delta_1 \delta_2 \rangle
 = \xi_{12} + \zeta_{012}
\end{equation}
which differs from the true correlation function $\xi_{12}$
by the three-point correlation function $\zeta_{012}$.
The three-point function in the real Universe appears well approximated by
the hierarchical model
$\zeta_{012} = Q ( \xi_{01} \xi_{12} + \xi_{02} \xi_{12} + \xi_{01} \xi_{02} )$
with $Q \approx 1$
(e.g.\ Fry \& Gazta\~naga 1994, table 8).
To ensure that the three-point term in equation (\ref{xiapp})
is small compared with the desired two-point term,
$\zeta_{012} \ll \xi_{12}$,
requires at least that $\xi_{01}$, $\xi_{02} \ll 1$.
Thus, on average, galaxy-bound observers
who want an unbiased measure of the correlation function
would be advised to exclude a local region around themselves
which is at least a correlation length in radius.

The importance of deleting the local region around the Milky Way needs
emphasizing.
If the local region is not deleted,
then the spherical distortion equation (\ref{xi})
will attempt to interpret the local overdensity as caused by
peculiar redshifts, which is clearly wrong.
The local region must be deleted in any case to ensure
that peculiar velocities are small compared with the depth of the structures
being observed, $v \ll r$, equation (\ref{v<r}).

\subsection{The peculiar velocity of the Milky Way}
\label{LGsec}

Formula (\ref{xi}) for the redshifted correlation function
is valid for randomly located observers in stationary frames of reference.
The Milky Way however is not stationary, but moving.
Happily,
this motion is rather accurately known from the dipole anisotropy of the CMB
(Kogut et al.\ 1993).
The linear part of the Milky Way's peculiar velocity is
the streaming motion ${\bmi V}$ of the Local Group (LG)
(Yahil, Tammann \& Sandage 1977)
with respect to the CMB frame.

In the LG frame, the peculiar redshift $s^{\rmn {LG}}$
of a galaxy at true position ${\bmi r}$ relative to the observer is
\begin{equation}
\label{sLG}
 s^{\rmn {LG}} = r + v - \hat{\bmi r} . {\bmi V} \ .
\end{equation}
Recapitulating the derivation of equation (\ref{bdelta}),
but starting from equation (\ref{sLG}) instead of (\ref{s}),
one concludes that,
to linear order,
the redshift space overdensity
$\sdeltaLG ({\bmi r})$
observed in the LG frame
differs from the
redshift space overdensity
$\sdelta ({\bmi r})$
in the CMB frame
by a dipole term directed along the LG motion ${\bmi V}$:
\begin{equation}
\label{bdeltaLG}
 \sdeltaLG ( {\bmi r} )
 = \sdelta ( {\bmi r} )
  + \alpha (r) {\hat{\bmi r} . {\bmi V} \over r}
 \ .
\end{equation}
The fact that the dipole term changes sign when $\alpha (r)$ goes negative,
that is, where the selection function $\Phi (r)$ is steeper than $r^{-2}$,
is the origin of the `rocket effect' noted by Kaiser (1987, section 2).
It follows from equation (\ref{bdeltaLG}) that
the redshift correlation function
$\sxiLG ( r_{12} , r_1 , r_2 )$
in the LG frame is related to that
$\sxi ( r_{12} , r_1 , r_2 )$
in the CMB frame
by
\begin{equation}
\label{xiLG}
 \sxiLG ( r_{12} , r_1 , r_2 ) =
  \sxi ( r_{12} , r_1 , r_2 )
  + \left[
   {\alpha ( r_1 ) V \sdeltaLG_1 ( r_1 \hat{\bmi V} ) \over 3 r_1}
   + {\alpha ( r_2 ) V \sdeltaLG_1 ( r_2 \hat{\bmi V} ) \over 3 r_2}
   + {\alpha ( r_1 ) \alpha ( r_2 ) V^2 \over 3 r_1 r_2}
  \right]
  \hat{\bmi r}_1 . \hat{\bmi r}_2
\end{equation}
where $\sdeltaLG_1 ( r \hat{\bmi V} )$
is the dipole component of the
redshift space overdensity in the LG frame
at depth $r$ in the direction of the LG motion $\bmi V$.
That is,
if the overdensity is expanded in spherical harmonics about the observer,
$\sdeltaLG ( {\bmi r} )
= \sum_{lm} \sdeltaLG_{lm} (r) Y_{lm} ( \hat{\bmi r} )$,
then $\sdeltaLG_1 ( {\bmi r} )
= \sum_m \sdeltaLG_{1m} (r) Y_{1m} ( \hat{\bmi r} )$
is the dipole component.
As in the case of the density,
the difference between the correlation functions in the two frames
is pure dipole.

There are basically three ways to deal with the
finite peculiar velocity $\bmi V$ of the Local Group with respect to
the CMB frame, when measuring redshift distortions.
The first is to work in the CMB frame using the known value of $\bmi V$.
The second is to treat $\bmi V$ as unknown,
and to average the right hand side of equation (\ref{xiLG})
over an ensemble of observers who measure structure
in frames at rest with respect to the local streaming frame.
The third way is to avoid the problem altogether by ignoring the dipole
component of the redshift correlation function.

Of these options the first,
to work in the CMB frame using the known value of $\bmi V$,
makes the most use of available information,
and is also the simplest,
which would seem to make it a clear winner.

To linear order in the supposedly small quantity $v / r$,
measuring $\sxi$ in the CMB frame is the same
as measuring $\sxiLG$ in the LG frame
and then using equation (\ref{xiLG}) to transform to the CMB frame.
The two measures do however differ in higher order.
Since for nearby structure the approximation
$v - \hat{\bmi r} . {\bmi V} \ll r$
is better than the approximation $v \ll r$,
it should be somewhat more accurate to measure $\sxiLG$
in the LG frame and then to transform to the CMB frame using
equation (\ref{xiLG}).
\section{The $\ddlnr$ Operator}

The symbol $\triangle$ is shorthand for
any two variables,
such as
($r_1 / r_{12}$ , $r_2 / r_{12}$),
which define the shape (angles)
of the triangle formed by the observer (= the Milky Way)
and the observed pair of galaxies
at positions ${\bmi r}_1$ and ${\bmi r}_2$ relative to the observer.

\subsection{The constancy of $\alpha$ in the distortion operator}
\label{alph}

The spherical distortion operator, the quantity in square brackets in
equation (\ref{bdelta}),
contains a term which depends on the function $\alpha ( r )$,
equation (\ref{alpha}),
which is the logarithmic slope of $r^2$ times the selection function
$\Phi (r)$ at depth $r$.
It is a basic assumption of this paper that $\alpha ( r )$
can be approximated by a constant
(or more generally by a function only of triangle shape $\triangle$),
to ensure that the operator
$\ddlnr$ considered in Section \ref{dlnr} below,
and also the logarithmic radial derivative operators considered in Section 4,
do in fact commute with the spherical distortion operator as claimed.

While $\alpha (r)$ is generally not constant in real redshift surveys,
it is typically a slowly varying function of depth $r$.
According to the uncertainty principle,
modes can be located in space within no better than a wavelength,
so the approximation of constant $\alpha$ might be expected
to be valid for modes whose wavelengths
are short compared with the scale over which $\alpha ( r )$ varies.

In a flux-limited redshift survey,
$\alpha$ is typically approximately constant at moderate depths,
a consequence of the power-law character of the luminosity function
at faint fluxes.
Since $\alpha$ decreases in importance for pairs at depths
greater than their separation,
the approximation of constant $\alpha$ should be valid
at least at separations
no larger than the depth at which $\alpha$ starts to
deviate significantly from a constant.

\subsection{The $\ddlnr$ operator}
\label{dlnr}

Let $\ddlnr$ denote the logarithmic derivative
with respect to pair separation $r_{12}$,
the shape (angles) of the triangle formed by
the observer and the observed pair of galaxies
being held fixed.

The operator
$\ddlnr$
possesses two key properties, discussed in the Introduction.
The first is that it commutes with the spherical distortion operator
(to the extent that $\alpha$ is constant or can be approximated
as a function of $\triangle$, Section \ref{alph}),
which implies that eigenfunctions of
$\ddlnr$
are also eigenfunctions of the spherical distortion operator.
The second property is that
the eigenfunctions of $\ddlnr$
form a complete set for
the (unredshifted) correlation function $\rxi (r_{12})$,
which is a function of separation $r_{12}$ alone.

A complete set of orthogonal eigenfunctions of $\ddlnr$ is
\begin{equation}
\label{phi}
 r_{12}^{- \gamma - {\rmn i} \omega_{12}}
\end{equation}
where $\omega_{12}$ takes all real values from negative to positive infinity,
and $\gamma$ is some fixed real number, chosen in practice to secure
convergence at $r_{12} \rightarrow 0$ and $r_{12} \rightarrow \infty$.
The choice of the units of the separation $r_{12}$ in equation (\ref{phi})
is a matter of convenience.
Let $\rxi_{\omega_{12}}$ denote
the representation of the real space correlation function $\rxi ( r_{12} )$
in the eigenfunctions $r_{12}^{- \gamma - {\rmn i} \omega_{12}}$:
\begin{equation}
\label{xir}
 \rxi (r_{12}) = \int_{- \infty}^{\infty}
  \rxi_{\omega_{12}} r_{12}^{- \gamma - {\rmn i} \omega_{12}} {\rmn d} \omega_{12}
\ \ \ , \ \ \ \ \ 
 \rxi_{\omega_{12}} =
  (2\pi )^{-1} \int_0^{\infty} \rxi ( r _{12} ) r_{12}^{\gamma + {\rmn i} \omega_{12}}
  {\rmn d} r_{12} / r_{12} \ .
\end{equation}
In effect, $\rxi_{\omega_{12}}$ is the Fourier transform of
$\rxi ( r_{12} ) r_{12}^\gamma$
with respect to logarithmic separation $\ln r_{12}$.
If the index $\gamma + {\rmn i} \omega_{12}$
were taken to be real rather than complex,
then $\rxi_{\omega_{12}}$ would be the Mellin transform of $\rxi (r_{12})$.
The reality of $\rxi (r_{12})$ implies the condition
\begin{equation}
 \rxi_{-\omega_{12}} = \rxi^{\ast}_{\omega_{12}} \;.
\end{equation}

Consider for example the case where the correlation function is
an exact power law $\rxi (r_{12}) = r_{12}^{-\gamma}$.
Then the correlation function in $\omega_{12}$-space is a Dirac delta-function,
$\rxi_{\omega_{12}} = \delta_{\rmn D} ( \omega_{12} )$,
provided that $\gamma$ in the expansion (\ref{xir})
is chosen equal to the actual power-law index $\gamma$
of the correlation function.

More realistically,
if the correlation function is flatter than $r_{12}^{-\gamma}$
at small scales and steeper than $r_{12}^{-\gamma}$
at larger scales,
then the correlation function in $\omega_{12}$-space should
look like a `window' about $\omega_{12} = 0$
of width $\sim 1 / \Delta \ln r$,
where $\Delta \ln r$ is roughly the logarithmic range of $r_{12}$
over which $\rxi$ approximates a power law $r_{12}^{-\gamma}$.

The representation $\sxi ( \omega_{12} , \triangle )$
in $\omega_{12}$-space of the redshift correlation function
is defined similarly to equation (\ref{xir}):
\begin{equation}
\label{bxir}
 \sxi ( r_{12} , \triangle ) = \int_{- \infty}^{\infty}
  \sxi ( \omega_{12} , \triangle ) r_{12}^{- \gamma - {\rmn i} \omega_{12}}
  {\rmn d} \omega_{12}
\ \ \ , \ \ \ \ \ 
 \sxi ( \omega_{12} , \triangle ) =
  (2\pi )^{-1} \int_0^{\infty} \sxi ( r_{12} , \triangle )
  r_{12}^{\gamma + {\rmn i} \omega_{12}}
  {\rmn d} r_{12} / r_{12} |_{\triangle} \ .
\end{equation}

\begin{figure}
\begin{center}
\leavevmode
\epsfbox{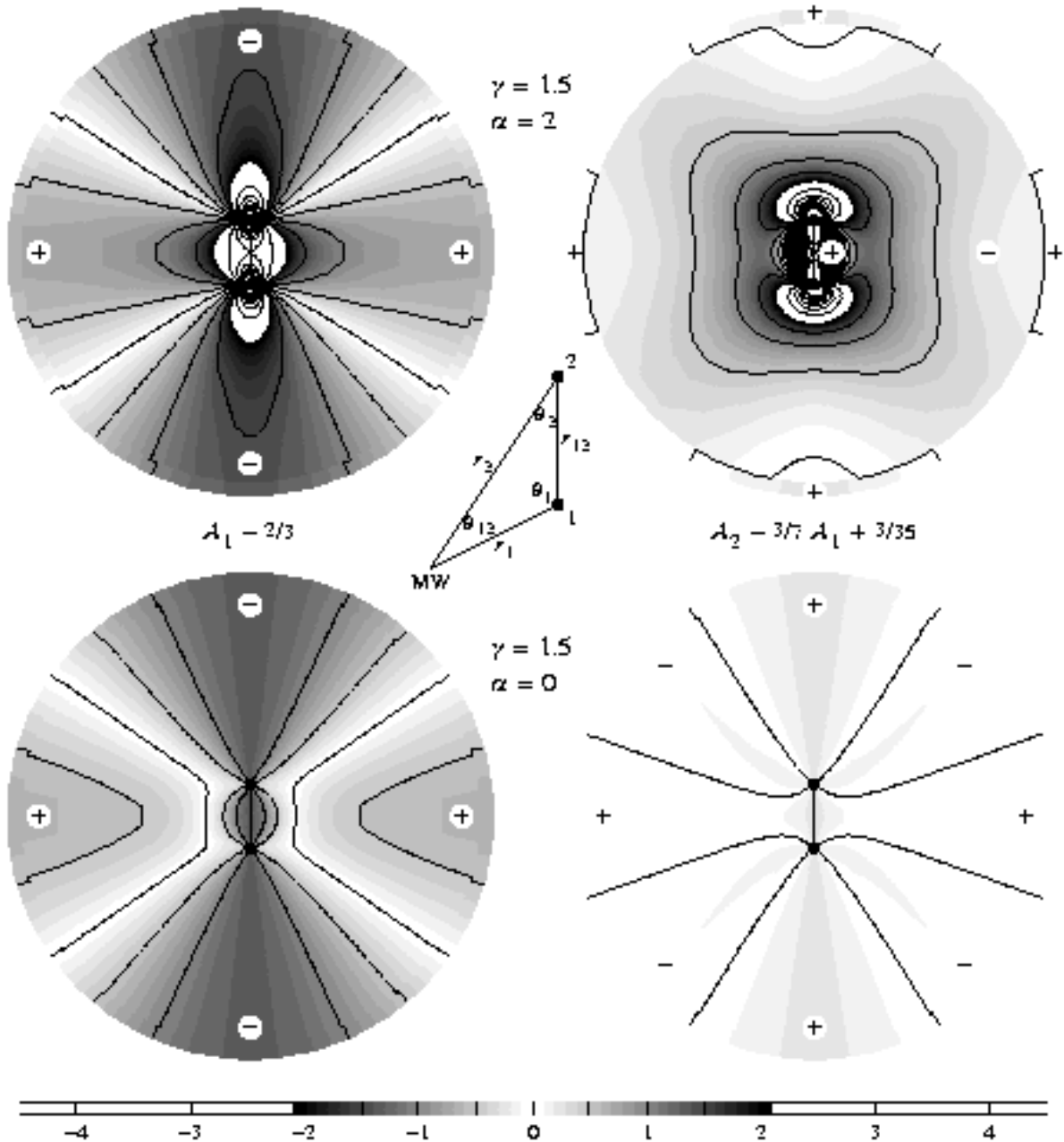}
\end{center}
\caption{
Contour diagrams showing, on the left, the `quadrupole'
amplitude
$A_1 - 2/3$,
and, on the right, the `hexadecapole'
amplitude
$A_2 - 3/7 A_1 + 3/35$,
equations (26), (27) and (38),
of the spherical distortion of the redshift correlation function
for the case of a power-law correlation function
$\rxi \propto r_{12}^{-\gamma}$ with index $\gamma = 1.5$.
To demonstrate the influence of the terms depending on
the shape of the selection function, i.e.\ on $\alpha$,
the top two contour diagrams are for $\alpha = 2$,
appropriate for a volume-limited sample,
while the bottom two are for $\alpha = 0$,
appropriate for a selection function $\Phi (r) \propto r^{-2}$.
The units are such that the unredshifted correlation function has
unit amplitude.
The geometry of the observer and the observed pair is illustrated
by the stick diagram at the centre.
In the diagrams, the observed pair,
represented by two large dots joined by a thick vertical line, is fixed,
while the position of the observer varies over the diagram.
Thus observers at 6 or 12 o'clock
observe pairs which are aligned along the line of sight,
while observers at 3 or 9 o'clock
observe pairs which are transverse to the line of sight.
The kink in the contours at the outermost radius in each diagram
is not an error;
the final ring shows the plane-parallel limit,
corresponding to infinite radius.
In the `quadrupole' distortion, on the left,
the correlation function appears reduced for pairs along the line of sight,
and enhanced for pairs transverse to the line of sight,
which is the expected squashing effect.
If one imagines the figures to be rotated about their vertical axes,
then volume elements in the resulting three-dimensional figure
are in correct proportion to the volume element
${\rmn d} \triangle = r_1 {\rmn d} r_1 r_2 {\rmn d} r_2 / r_{12}^4$
of the geometrical configuration.
Greyscales are graduated at intervals of $0.1$ in the interval $[ - 2 , 2]$,
with contours at intervals of $0.5$.
Outside $[ - 2 , 2 ]$, contours are at intervals of 1, up to $\pm 10$.
}
\label{fig}
\end{figure}

In $\omega_{12}$-space,
the spherical distortion equation (\ref{xi}) becomes
\begin{equation}
\label{bxiomega}
 \sxi ( \omega_{12} , \triangle ) =
  [ 1 + \ff A_1 (\omega_{12} , \triangle )
   + \ff^2 A_2 (\omega_{12} , \triangle ) ]
  \, \rxi_{\omega_{12}}
\end{equation}
where, with
\begin{equation}
 \eta_{12} \equiv \gamma + {\rmn i} \omega_{12}
 \ ,
\end{equation}
the coefficient of $\ff$ is
\begin{equation}
\label{A1}
 A_1 (\omega_{12} , \triangle ) =
  {2 \over 3}
  - {4 \eta_{12} \over 3 ( 3 - \eta_{12} )} B_2 ( \triangle )
  + {\alpha \over ( 3 - \eta_{12} )} B_1 ( \triangle )
\end{equation}
while the coefficient of $\ff^2$ is
\[
 A_2 (\omega_{12} , \triangle ) =
  {1 \over 5}
  - {4 \eta_{12} \over 7 ( 3 - \eta_{12} )} B_2 ( \triangle )
  + {8 \eta_{12} ( 2 + \eta_{12} ) \over 35 ( 3 - \eta_{12} ) ( 5 - \eta_{12} )}
   B_4 ( \triangle )
  - {2 ( 1 - \eta_{12} ) \over ( 3 - \eta_{12} ) ( 5 - \eta_{12} )}
   B_3 ( \triangle )
\]
\begin{equation}
\label{A2}
\ \ \ \ \ \ \ \ \ \ \ \ \ \ \ \ \ \ 
  + {\alpha \over 5 - \eta_{12}} \left( 1 + {\alpha \over 2 - \eta_{12}} \right)
   \left[
   B_1 ( \triangle )
   - \left( {4 - \eta_{12} \over 3 - \eta_{12}} \right) B_3 ( \triangle )
   \right]
   \ .
\end{equation}
The shape-functions $B_i ( \triangle )$
in equations (\ref{A1}) and (\ref{A2})
are (with a constant function $B_0$ to complete the set)
\begin{equation}
\label{B0}
 B_0 ( \triangle ) \equiv
  1
\end{equation}
\begin{equation}
\label{B1}
 B_1 ( \triangle ) \equiv
  {r_{12} \mu_1 \over r_1} + {r_{12} \mu_2 \over r_2}
  = {r_{12}^2 \mu_1 \mu_2 \over r_1 r_2} + 1 - \mu_{12}^2
  = - \mbox{} {r_{12}^2 \mu_{12} \over r_1 r_2} + 2 ( 1 - \mu_{12}^2 )
\end{equation}
\begin{equation}
\label{B2}
 B_2 ( \triangle ) \equiv
  {1 \over 2} \left[ P_2 ( \mu_1 ) + P_2 ( \mu_2 ) \right]
\end{equation}
\begin{equation}
\label{B3}
 B_3 ( \triangle ) \equiv
  1 - \mu_{12}^2
\end{equation}
\begin{equation}
\label{B4}
 B_4 ( \triangle ) \equiv
  {1 \over 8} ( 35 \mu_1^2 \mu_2^2 - 15 \mu_1^2 - 15 \mu_2^2 + 3 )
\end{equation}
where
$P_2 ( \mu ) = \frac{3}{2} \mu^2 - \frac{1}{2}$
is the quadrupole Legendre polynomial,
and
$\mu_{12} \equiv \cos \theta_{12}$,
$\mu_1 \equiv \cos \theta_1$
and $\mu_2 \equiv \cos \theta_2$
are the cosines of the interior angles of the triangle formed
by the observer and the observed pair of galaxies,
as illustrated in Fig.\ \ref{fig}:
\begin{equation}
 \mu_{12}
  = {r_1^2 + r_2^2 - r_{12}^2 \over 2 r_1 r_2}
 \ \ \ , \ \ \ \ \ 
 \mu_1
  = {r_{12}^2 + r_1^2 - r_2^2 \over 2 r_{12} r_1}
 \ \ \ , \ \ \ \ \ 
 \mu_2
  = {r_{12}^2 + r_2^2 - r_1^2 \over 2 r_{12} r_2}
 \ .
\end{equation}

The plane-parallel limit
is attained when the distance to each galaxy of a pair is large
compared with their separation, $r_1$, $r_2 \gg r_{12}$.
Here $\mu_1 \rightarrow - \mu_2$ and $\mu_{12} \rightarrow 1$,
the functions $B_1$ and $B_3$ vanish,
and the functions $B_2$ and $B_4$ go over to Legendre polynomials
$B_2 ( \triangle ) \rightarrow P_2 ( \mu )$ and
$B_4 ( \triangle ) \rightarrow P_4 ( \mu )$
in the cosine $\mu \approx | \mu_1 | \approx | \mu_2 |$
of the angle to the line of sight.
Thus the spherical distortion coefficients $A_1$ and $A_2$
in equations (\ref{A1}) and (\ref{A2}) go over to their plane-parallel limits
(compare Hamilton 1992, equations [10]--[12])
\begin{equation}
A_1 \rightarrow
\label{A1far}
  {2 \over 3}
  - {4 \eta_{12} \over 3 ( 3 - \eta_{12} )}
   P_2 ( \mu )
\end{equation}
\begin{equation}
\label{A2far}
 A_2 \rightarrow
  {1 \over 5}
  - {4 \eta_{12} \over 7 ( 3 - \eta_{12} )}
   P_2 ( \mu )
  + {8 \eta_{12} ( 2 + \eta_{12} ) \over 35 ( 3 - \eta_{12} ) ( 5 - \eta_{12} )}
   P_4 ( \mu )
 \ .
\end{equation}
In the plane-parallel limit,
the distortion decomposes naturally into
a sum of mutually orthogonal parts,
the harmonics $P_l ( \mu )$,
but in the general case the shape-functions $B_i ( \triangle )$,
equations (\ref{B0})--(\ref{B4}), are not orthogonal.

In the opposite limit where the distance to one of the galaxies, say 1,
of a pair is much smaller than their separation, $r_1 \ll r_{12}$,
then $r_2 \rightarrow r_{12}$, $\mu_{12} \rightarrow - \mu_{1}$
and $\mu_2 \rightarrow 1$,
the $B_1$ terms become large compared with the others,
and equations (\ref{A1}) and (\ref{A2}) reduce to
\begin{equation}
\label{A1near}
 A_1 \rightarrow
  {\alpha \over ( 3 - \eta_{12} )}
   {r_{12} \mu_1 \over r_1}
\end{equation}
\begin{equation}
\label{A2near}
 A_2 \rightarrow
  {\alpha \over 5 - \eta_{12}} \left( 1 + {\alpha \over 2 - \eta_{12}} \right)
   {r_{12} \mu_1 \over r_1} \ .
\end{equation}
Here the distortion is dominated by a dipole about the near galaxy,
reflecting the streaming motion of galaxies past the stationary observer.
The divergence of the distortion as $r_1 \rightarrow 0$
is an artefact
which results from a violation of the presumption (\ref{v<r}) that $v \ll r_1$.
The divergence is removed in practice by deleting the local region.
Note that the validity of the linear approximation requires
not that the dipole term be small, but only the weaker constraint
that $\rxi (r_{12}) r_{12}/r_1$ be small.

\subsection{Picture}

Fig.\ \ref{fig} illustrates the
spherical distortion of the redshift correlation function
for the case of a pure power law $\rxi \propto r_{12}^{-\gamma}$
with index $\gamma = 1.5$.
As mentioned above,
$\rxi_{\omega_{12}}$ is a Dirac delta-function at $\omega_{12} = 0$
for a pure power law,
so $\eta_{12} = \gamma = 1.5$ in equations (\ref{A1}) and (\ref{A2})
for this case.
Fig.\ \ref{fig} shows not the spherical distortion
coefficients $A_1$ and $A_2$
directly,
but rather the combinations
$A_1 - 2/3$
and $A_2 - 3/7 A_1 + 3/35$
which go over respectively to pure quadrupole and pure hexadecapole
distortions in the plane-parallel limit
(cf.\ equations [\ref{A1far}] and [\ref{A2far}]).
In terms of these coefficients,
which we refer to loosely as `quadrupole' and `hexadecapole' coefficients,
the distortion equation (\ref{bxiomega}) is
(the normalization of the coefficients here differs
by factors of $3/4$ and $35/8$ from the normalization
of the `quadrupole' and `hexadecapole' correlation functions
$\sXi_2$ and $\sXi_4$ of equations [\ref{Xi2}] \& [\ref{Xi4}]
in Section \ref{rrep})
\begin{equation}
\label{xiquad}
 \sxi =
  \left[ \left( 1 + \frac{2}{3} \ff + \frac{1}{5} \ff^2 \right)
  + \left( \ff + \frac{3}{7} \ff^2 \right) \left( A_1 - \frac{2}{3} \right)
  + \ff^2 \left( A_2 - \frac{3}{7} A_1 + \frac{3}{35} \right) \right] \, \rxi
 \ .
\end{equation}
To demonstrate the influence of the terms depending on
the shape of the selection function, i.e.\ on $\alpha$,
equation (\ref{alpha}),
Fig.\ \ref{fig} shows two cases,
the first with $\alpha = 2$, appropriate for a volume-limited sample,
the second with $\alpha = 0$,
appropriate for a selection function $\Phi (r) \propto r^{-2}$.

One of the striking aspects of Fig.\ \ref{fig}
is how large the distortions become for `nearby' pairs,
those whose distances are less than or
comparable to their separation, $r_1$, $r_2 \la r_{12}$
(note that such pairs need not be physically nearby if the separation $r_{12}$
is large).
The large nearby distortion is produced by the `dipole'
term $B_1 ( \triangle )$, equation (\ref{B1}).
This dipole should not be confused with the dipole in the correlation function,
whose general form is $\mu_{12}$ times some function of $r_1$
and $r_2$, independent of separation $r_{12}$.

The amplitude of the nearby dipole distortion depends on $\alpha$.
This raises the concern that approximating $\alpha$ by a constant
[or more generally by a function $\alpha ( \triangle )$]
may be unreasonable,
if in fact the amplitude of the distortion depends sensitively
on $\alpha$.
However,
the problem is not as bad as it seems,
because the functions $B_i$ for $i \neq 1$
are all orthogonal to $B_1$ in the limit where one galaxy of a pair
is much closer than their separation.
That is,
$B_i$ for $i \neq 1$ all go over to sums of a constant and a quadrupole
in the limit $r_1 / r_{12}$ or $r_2 / r_{12} \rightarrow 0$,
so are orthogonal to the diverging dipole.
Thus uncertainty caused by approximating $\alpha$
translates mainly into uncertainty about the amplitude
of the $B_1$ part of the distortion,
with lesser impact on the other parts.

It is to be noted that the spherical distortion
coefficients $A_1$ and $A_2$,
equations (\ref{A1}) and (\ref{A2}),
depend on $\alpha$ only through the shape-functions $B_1$ and $B_3$.
As will be seen in Section \ref{rrep},
this has the consequence that the spherical distortion separates
naturally into a part
($\sXi_1$ and $\sXi_3$, equations [\ref{Xi1}] and [\ref{Xi3}])
that depends on $\alpha$,
and a part
($\sXi_0$, $\sXi_2$ and $\sXi_4$, equations [\ref{Xi0}]--[\ref{Xi4}])
that has no explicit dependence on $\alpha$.
\section{Logarithmic Spherical Waves}
\label{waves}

The trick in Section 3 of expanding the
correlation function in eigenfunctions of a single operator
fails in Fourier space.
For example,
one would naturally think to expand the (unredshifted) power spectrum $P ( k )$,
which is the Fourier transform of the correlation function $\rxi ( r_{12} )$,
in eigenfunctions of an operator $\partial / \partial \ln k$.
The procedure fails because the power spectrum is correctly a function
$\delta_{\rmn D} ( {\bmi k}_1 + {\bmi k}_2 ) P ( k_1 )$
of two arguments, not one.
The Dirac delta-function $\delta_{\rmn D} ( {\bmi k}_1 + {\bmi k}_2 )$
arises from the statistical homogeneity,
or translational invariance, of clustering in space,
the Fourier modes $\tilde\rdelta ( {\bmi k} )$
being by definition the eigenfunctions of the translation operator
${\rmn i} \bnabla$.
In redshift space radial distortions destroy homogeneity,
although statistical isotropy about the observer is preserved.
Thus the delta-function $\delta_{\rmn D} ( {\bmi k}_1 + {\bmi k}_2 )$
cannot be an eigenfunction of the spherical distortion operator,
and it cannot be factored out.
The situation here contrasts with the plane-parallel limit,
where translation symmetry is preserved in redshift space,
and the delta-function factors out of the distortion equation.

So is there a basis of eigenfunctions for spherical distortions
analogous to the Fourier modes for plane-parallel distortions?
As just mentioned, Fourier modes themselves fail,
precisely because spherical distortions destroy homogeneity.

Now the great advantage of the power spectrum is that,
for Gaussian fluctuations,
the (unredshifted) modes $\tilde\rdelta ( {\bmi k} )$ are independent.
Their independence is a consequence of homogeneity
coupled with the Gaussian assumption.
Homogeneity implies that modes with different wavevectors $\bmi k$
are uncorrelated,
$\langle \rdelta ( {\bmi k}_1 ) \rdeltaast \! ( {\bmi k}_2 ) \rangle = 0$
for ${\bmi k}_1 \neq {\bmi k}_2$,
while Gaussianity means that all higher order correlations vanish.
Linear fluctuations may well be Gaussian,
a consequence of an alliance between the Central Limit Theorem
and physical smoothing processes in the early Universe.

Spherical distortions destroy homogeneity,
but they preserve isotropy about the observer.
Thus the advantage of mode independence is preserved to the greatest
extent if the density is expanded in eigenfunctions
of the generator of rotations,
which is the angular momentum operator ${\bmi L}$
(e.g.\ Landau \& Lifshitz 1958).
These eigenfunctions are spherical harmonics,
and symmetry about the observer implies that spherical harmonic modes
$\rdelta_{lm} ( k )$ with different $lm$
will be independent for Gaussian fluctuations.
This basic advantage of the spherical harmonic
modes was exploited by Fisher et al.\ (1994b),
and by Heavens \& Taylor (1995).

This leaves the problem of finding radial eigenmodes
for the spherical distortion operator.
Here one notices that,
at least to the extent that the slowly varying function $\alpha ( r )$
is constant (see Section \ref{alph}),
the spherical distortion operator in equation (\ref{bdelta}) is scale-free.
A consequence of this is that the logarithmic derivative
$\partial / \partial \ln r$
with respect to depth $r$
(or alternatively the logarithmic derivative
$\partial / \partial \ln k$
with respect to radial wavevector $k$)
commutes with the spherical distortion operator,
and its eigenfunctions therefore provide a basis of radial eigenmodes.

We proceed with this idea further in Section \ref{complete},
but first it is necessary to write down some standard definitions.

\subsection{Fourier space definitions}
\label{fourier}

Let $\tilde\sdelta ( {\bmi k} )$
denote the Fourier transform of the overdensity:
\begin{equation}
 \tilde\sdelta ( {\bmi k} ) =
  ( 2 \pi )^{-3/2} \int \sdelta ( {\bmi r} )
  e^{{\rmn i} {\bmi k} . {\bmi r}} {\rmn d}^3 \! r
\ \ \ , \ \ \ \ \ 
 \sdelta ( {\bmi r} ) =
  ( 2 \pi )^{-3/2} \int \tilde\sdelta ( {\bmi k} )
  e^{- {\rmn i} {\bmi k} . {\bmi r}} {\rmn d}^3 \! k \ .
\end{equation}

The (unredshifted) power spectrum $P ( k )$
is by definition the Fourier transform of the correlation function
$\rxi ( r_{12} )$,
with the conventional normalization
\begin{equation}
\label{Pk}
 P ( k ) =
  \int \rxi ( r_{12} )
  e^{{\rmn i} {\bmi k} . {\bmi r_{12}}} {\rmn d}^3 \! r_{12}
\ \ \ , \ \ \ \ \ 
 \rxi ( r_{12} ) =
  \langle \rdelta ( {\bmi r}_1 ) \rdelta ( {\bmi r}_2 ) \rangle =
  ( 2 \pi )^{-3} \int P ( k )
  e^{- {\rmn i} {\bmi k} . {\bmi r_{12}}} {\rmn d}^3 \! k \ .
\end{equation}

In redshift space,
it is necessary to retain the dependence of the correlation function
$\sxi ( {\bmi r}_1 , {\bmi r}_2 )$
and its Fourier transform the power spectrum
$\tilde\sxi ( {\bmi k}_1 , {\bmi k}_2 )$
on both their arguments:
\begin{equation} 
\label{kk}
 \tilde\sxi ( {\bmi k}_1 , {\bmi k}_2 ) =
  \langle \tilde\sdelta ( {\bmi k}_1 ) \tilde\sdelta ( {\bmi k}_2 ) \rangle =
  ( 2 \pi )^{-3} \int
  \sxi ( {\bmi r}_1 , {\bmi r}_2 )
  e^{{\rmn i} ( {\bmi k}_1 . {\bmi r}_1 + {\bmi k}_2 . {\bmi r}_2 )}
  {\rmn d}^3 \! r_1 {\rmn d}^3 \! r_2
\end{equation}
\begin{equation} 
\label{rr}
 \sxi ( {\bmi r}_1 , {\bmi r}_2 ) =
  \langle \sdelta ( {\bmi r}_1 ) \sdelta ( {\bmi r}_2 ) \rangle =
  ( 2 \pi )^{-3} \int
  \tilde\sxi ( {\bmi k}_1 , {\bmi k}_2 )
  e^{- {\rmn i} ( {\bmi k}_1 . {\bmi r}_1 + {\bmi k}_2 . {\bmi r}_2 )}
  {\rmn d}^3 \! k_1 {\rmn d}^3 \! k_2 \ .
\end{equation}
The unredshifted power spectrum
$\tilde\rxi ( {\bmi k}_1 , {\bmi k}_2 )$
is related to the `reduced' power spectrum $P ( k )$, equation (\ref{Pk}),
by
\begin{equation}
\label{Pk2}
 \tilde\rxi ( {\bmi k}_1 , {\bmi k}_2 ) =
  \delta_{\rmn D} ( {\bmi k}_1 + {\bmi k}_2 ) P ( k_1 ) \ .
\end{equation}

\subsection{A complete set of commuting operators}
\label{complete}

In Section \ref{SDO} it was shown that, for linear fluctuations,
the overdensity $\sdelta ( {\bmi r} )$ in redshift space
is described by a spherical distortion operator acting on
the real overdensity $\rdelta ( {\bmi r} )$, equation (\ref{bdelta}).
A complete set of commuting operators
(to the extent that the quantity $\alpha$ is constant, Section \ref{alph})
for the spherical distortion operator is
\begin{equation}
\label{op}
 {\partial \over \partial \ln r} = - {\partial \over \partial \ln k} - 3
 \ ,\ \ 
 L^2
 \ ,\ \ 
 L_z
 \ .
\end{equation}
Here $\partial / \partial \ln r$ is
the logarithmic derivative with respect to depth $r$,
which is the same, up to a change of sign and a constant,
as the logarithmic derivative
$\partial / \partial \ln k$
with respect to radial wavevector $k$ in Fourier space.
The operators $L^2$ and $L_z$
are the square and $z$-component
(along some arbitrary axis) of the angular momentum operator
${\bmi L} = {\rmn i} {\bmi r} \times \partial / \partial {\bmi r}
= {\rmn i} {\bmi k} \times \partial / \partial {\bmi k}$,
which is the same operator in real and Fourier space.
The eigenfunctions of $\partial / \partial \ln r$ or
$\partial / \partial \ln k$
are radial waves in logarithmic depth $r$ or wavevector $k$,
while the eigenfunctions of $L^2$ and $L_z$ are the usual orthonormal
spherical harmonics $Y_{lm}$.
Thus the eigenfunctions of the commuting set (\ref{op})
are spherical waves with radial parts in logarithmic real or Fourier space.

Equations (\ref{olmr})--(\ref{C}) below are valid for both redshifted
and unredshifted correlation functions.
Let $\sdelta_{\omega lm}$ denote the representation of
the overdensity $\sdelta ( {\bmi r} )$
as spherical waves in logarithmic real space:
\begin{equation}
\label{olmr}
 \sdelta ( {\bmi r} ) =
  ( 2 \pi )^{-\half} \int_{-\infty}^{\infty} \sum_{lm}
  \sdelta_{\omega lm} r^{- \gamma /2 - {\rmn i} \omega}
  Y_{lm} ( \hat{\bmi r} ) {\rmn d} \omega
\ \ \ , \ \ \ \ \ 
 \sdelta_{\omega lm} =
  ( 2 \pi )^{-\half} \int
  \sdelta ( {\bmi r} ) r^{- 3 + \gamma /2 + {\rmn i} \omega}
  Y^{\ast}_{lm} ( \hat{\bmi r} ) {\rmn d}^3 \! r
\end{equation}
and let
$\tilde\sdelta_{\omega lm}$
denote the alternative representation
of the overdensity as spherical waves in logarithmic Fourier space
\begin{equation}
\label{olmk}
 \tilde\sdelta ( {\bmi k} ) =
  ( 2 \pi )^{-\half} \int_{-\infty}^{\infty} \sum_{lm}
  \tilde\sdelta_{\omega lm} k^{- 3 + \gamma /2 + {\rmn i} \omega}
  Y_{lm} ( \hat{\bmi k} ) {\rmn d} \omega
\ \ \ , \ \ \ \ \ 
 \tilde\sdelta_{\omega lm} =
  ( 2 \pi )^{-\half} \int
  \tilde\sdelta ( {\bmi k} ) k^{- \gamma /2 - {\rmn i} \omega}
  Y^{\ast}_{lm} ( \hat{\bmi k} ) {\rmn d}^3 \! k \ .
\end{equation}
The index $\gamma$ in equations (\ref{olmr}) and (\ref{olmk})
is the same $\gamma$ as in the representation (\ref{xir}) or (\ref{bxir})
of the correlation function in eigenfunctions
$r_{12}^{-\gamma - {\rmn i} \omega_{12}}$.
The factor of a half which multiplies the index $\gamma$
in equations (\ref{olmr}) and (\ref{olmk})
arises because the correlation function is the square of the density.
The reality conditions $\sdeltaast \! ( {\bmi r} ) = \sdelta ( {\bmi r} )$,
hence $\tilde\sdeltaast \! ( {\bmi k} ) = \tilde\sdelta ( - {\bmi k} )$,
along with the usual properties
$Y^{\ast}_{lm} = (-)^m Y_{l , -m}$
and $Y_{lm} ( - \hat{\bmi k} ) = (-)^l Y_{lm} ( \hat{\bmi k} )$
of the spherical harmonics,
imply
\begin{equation}
\label{olmreal}
 \sdeltaast_{\omega lm} = (-)^m \sdelta_{-\omega , l , -m}
\ \ \ , \ \ \ \ \ 
 \tilde\sdeltaast_{\omega lm} = (-)^{l+m} \tilde\sdelta_{-\omega , l , -m}
 \ .
\end{equation}
Equating
the expansion of $\sdelta ( {\bmi r} )$ in equation (\ref{olmr})
to the Fourier transform of
the expansion of $\tilde\sdelta ( {\bmi k} )$ in equation (\ref{olmk})
shows that
the eigenfunctions $\sdelta_{\omega lm}$ and $\tilde\sdelta_{\omega lm}$
are the same up to factors $C ( \gamma /2 \!+\! {\rmn i} \omega , l )$:
\begin{equation}
\label{C}
 \sdelta_{\omega lm} =
  C ( \gamma /2 \!+\! {\rmn i} \omega , l )
  \, \tilde\sdelta_{\omega lm}
 \ \ \ , \ \ \ \ \ 
 C ( \eta , l ) \equiv
  (-{\rmn i})^l 2^{- ( \oneandahalf ) + \eta}
  {\Gamma \left[ {\eta + l \over 2} \right] \over
   \Gamma \left[ {3 - \eta + l \over 2} \right]} \ .
\end{equation}
The $\Gamma$ functions come from
integrals of power laws with spherical Bessel functions.
The two $\Gamma$ functions in equation (\ref{C})
can be reduced to the product of a single $\Gamma$ function with
a sine function and a rational function,
but the expression (\ref{C}) manifests the symmetry between
the two representations $\sdelta_{\omega lm}$ and $\tilde\sdelta_{\omega lm}$.

In $\omega lm$-space,
with
\begin{equation}
 \eta \equiv \gamma / 2 + {\rmn i} \omega
\end{equation}
the spherical distortion equation (\ref{bdelta}) becomes
\begin{equation}
\label{bdeltao}
 \sdelta_{\omega l m} =
  \left[ 1 + \ff {( 2 - \eta ) ( 1 - \eta + \alpha )
   \over ( 2 - \eta ) ( 3 - \eta ) - l ( l + 1 )} \right]
  \rdelta_{\omega l m}
\end{equation}
with an identical relation between $\tilde\sdelta_{\omega l m}$
and $\tilde\rdelta_{\omega l m}$.
As expected,
the spherical distortion operator reduces to eigenvalues in $\omega lm$-space.

\subsection{The correlation function of logarithmic spherical waves}

As will be seen shortly,
the normalization of the `Fourier' representation
$\tilde\sdelta_{\omega l m}$, equation (\ref{olmk}),
leads to simpler expressions for the correlation function
in $\omega lm$-space
than that of the `real' representation
$\sdelta_{\omega l m}$, equation (\ref{olmr}).
We therefore particularize to the former.

Statistical isotropy about the observer implies that
the correlation function of logarithmic spherical waves
$\tilde\sdelta_{\omega l m}$
is diagonal in the angular indexes $lm$:
\begin{equation}
\label{xiolm}
 \langle \tilde\sdelta_{\omega_1 l_1 m_1}
  \tilde\sdelta_{\omega_2 l_2 m_2} \rangle
  = ( - )^{l_1 + m_1}
  \rdelta_{l_1 l_2} \, \rdelta_{m_1 , - m_2} \,
  \tilde\sxi_{\omega_1 \omega_2 l_1}
\end{equation}
where the $\rdelta_{l_1 l_2}$ and $\rdelta_{m_1 , - m_2}$
denote Kronecker deltas, not to be confused with the overdensity.
The extraneous minus signs in equation (\ref{xiolm})
arise from pair exchange symmetry and reality,
as in equation (\ref{olmreal}),
and would disappear if the correlation function in equation (\ref{xiolm})
were defined by
$\langle \tilde\sdelta \tilde\sdeltaast \rangle$
rather than
$\langle \tilde\sdelta \tilde\sdelta \rangle$.
Equation (\ref{xiolm}) is valid for both redshifted and unredshifted
correlation functions.

The unredshifted correlation function
$\tilde\rxi_{\omega_1 \omega_2 l}$
is, from equations (\ref{kk}), (\ref{Pk2}), (\ref{olmk}), and (\ref{xiolm}),
\[ 
 \tilde\rxi_{\omega_1 \omega_2 l} =
  (-)^{l+m}
  \langle \tilde\rdelta_{\omega_1 l m} \tilde\rdelta_{\omega_2 , l , -m} \rangle
\] 
\begin{equation} 
\ \ \ \ \ \ 
\label{xiol}
  =
  ( 2 \pi )^{-1} \int
  \delta_{\rmn D} ( {\bmi k}_1 + {\bmi k}_2 ) P ( k_1 )
  k_1^{- \gamma /2 - {\rmn i} \omega_1}
  k_2^{- \gamma /2 - {\rmn i} \omega_2}
  (-)^l
  Y_{lm} ( \hat{\bmi k}_1 )
  Y^{\ast}_{lm} ( \hat{\bmi k}_2 )
  {\rmn d}^3 \! k_1
  {\rmn d}^3 \! k_2
  \ .
\end{equation}
Equation (\ref{xiol}) reduces to the simple result
that the unredshifted correlation function $\tilde\rxi_{\omega_1 \omega_2 l}$
is a function only of the sum
$\omega_1 + \omega_2$:
\begin{equation}
\label{xikred}
 \tilde\rxi_{\omega_1 \omega_2 l} = \tilde\rxi_{\omega_{12}}
 \ ,\ \ 
 \omega_{12} = {\omega_1 + \omega_2}
\end{equation}
where the reduced correlation function $\tilde\rxi_{\omega_{12}}$
in equation (\ref{xikred}) is
the representation of the power spectrum $P ( k )$
in eigenfunctions $k^{-3 + \gamma + {\rmn i} \omega_{12}}$
\begin{equation}
\label{xio}
 P ( k ) = \int_{- \infty}^{\infty}
  \tilde\rxi_{\omega_{12}}
  k^{-3 + \gamma + {\rmn i} \omega_{12}} {\rmn d} \omega_{12}
\ \ \ , \ \ \ \ \
 \tilde\rxi_{\omega_{12}} =
  ( 2 \pi )^{-1} \int_0^{\infty} P ( k )
  k^{3 - \gamma - {\rmn i} \omega_{12}} {\rmn d} k / k \ .
\end{equation}
Equation (\ref{xio}) is entirely analogous to the representation
(\ref{xir}) of the correlation function in
eigenfunctions $r_{12}^{- \gamma - {\rmn i} \omega_{12}}$.
Equating
the expansion of $\rxi ( r_{12} )$ in equation (\ref{xir})
to the Fourier transform of
the expansion of $P ( k )$ in equation (\ref{xio})
shows that the `real' $\rxi_{\omega_{12}}$
and `Fourier' $\tilde\rxi_{\omega_{12}}$
unredshifted correlation functions differ only by a factor
\begin{equation}
\label{xioC}
 \rxi_{\omega_{12}} = ( 2 \pi )^{-\oneandahalf} C ( \gamma + {\rmn i} \omega_{12} , 0 )
  \, \tilde\rxi_{\omega_{12}}
\end{equation}
where $C$ is the function defined in equation (\ref{C}).

It follows from equations (\ref{C}), (\ref{xikred}) and (\ref{xioC})
that the unredshifted correlation function
$\rxi_{\omega_1 \omega_2 l}$
of spherical waves in logarithmic real space
is related to the unredshifted correlation function $\rxi_{\omega_{12}}$
defined in equation (\ref{xir}) by
\begin{equation}
\label{xired}
 \rxi_{\omega_1 \omega_2 l} =
  (-)^m \langle \rdelta_{\omega_1 l m} \rdelta_{\omega_2 , l , -m} \rangle =
  (-)^l ( 2 \pi )^{\oneandahalf}
  {C ( \gamma /2 + {\rmn i} \omega_1 , l ) \, C ( \gamma /2 + {\rmn i} \omega_2 , l )
  \over C ( \gamma + {\rmn i} \omega_{12} , 0 )}
  \, \rxi_{\omega_{12}}
\end{equation}
which is more complicated than the corresponding simple `Fourier' relation
(\ref{xikred}).

The equality $\omega_{12} = \omega_1 + \omega_2$
is a consequence of the equality of operators
\begin{equation}
\label{expansion}
 \left. {\partial \over \partial \ln r_{12}} \right|_{\triangle}
  = {\partial \over \partial \ln r_1} + {\partial \over \partial \ln r_2}
 \ .
\end{equation}
Equation (\ref{expansion}) can be interpreted as meaning that
an expansion (by a factor) of the separation $r_{12}$ at fixed triangle shape
is equivalent to a combined expansion of the legs $r_1$ and $r_2$
at fixed angle between them.
An equivalent statement is valid in Fourier space.

In $\omega lm$-space,
with
\begin{equation}
 \eta_1 \equiv \gamma / 2 + {\rmn i} \omega_1
 \ \ \ ,\ \ \ \ \ 
 \eta_2 \equiv \gamma / 2 + {\rmn i} \omega_2
\end{equation}
the spherical distortion equation (\ref{xi}) becomes
\[ 
 \tilde\sxi_{\omega_1 \omega_2 l} =
  (-)^{l+m}
  \langle \tilde\sdelta_{\omega_1 l m} \tilde\sdelta_{\omega_2 , l , -m} \rangle
\] 
\begin{equation} 
\ \ \ \ \ 
\label{bxiol}
  =
  \left[ 1 + \ff {( 2 - \eta_1 ) ( 1 - \eta_1 + \alpha )
   \over ( 2 - \eta_1 ) ( 3 - \eta_1 ) - l ( l + 1 )} \right]
  \left[ 1 + \ff {( 2 - \eta_2 ) ( 1 - \eta_2 + \alpha )
   \over ( 2 - \eta_2 ) ( 3 - \eta_2 ) - l ( l + 1 )} \right]
  \tilde\rxi_{\omega_{12}}
  \ .
\end{equation}

In the plane-parallel limit,
$\omega_1 \approx -\omega_2 \rightarrow \infty$ and $l \rightarrow \infty$,
and the spherical distortion equation (\ref{bxiol}) reduces to
\begin{equation}
\label{bxiolpar}
 \tilde\sxi_{\omega_1 \omega_2 l} =
  ( 1 + \ff \mu^2 )^2
  \tilde\rxi_{\omega_{12}}
\ \ \ , \ \ \ \ \ 
 \mu^2
  \approx {\omega_1^2 \over \omega_1^2 + l^2}
  \approx {\omega_2^2 \over \omega_2^2 + l^2}
\end{equation}
which is Kaiser's equation (\ref{Kaiser}),
except that the power spectrum is expressed
in the representation $\tilde\rxi_{\omega_{12}}$, equation (\ref{xio}).
\section{Measuring Omega}

\subsection{Considerations of optimality}

The spherical distortion equation (\ref{bxiomega})
in the $\omega_{12} \triangle$-representation
is a linear equation in the unknown quantities
$\ff^i \rxi_{\omega_{12}}$:
\begin{equation}
\label{xisum}
 \sxi ( \omega_{12} , \triangle )
  =
 A_i ( \omega_{12} , \triangle ) \ff^i \rxi_{\omega_{12}}
\end{equation}
where indices $i$ run over 0, 1, 2,
and we adopt the summation convention for these indices.
The spherical distortion equation (\ref{bxiol})
in the $\omega lm$-representation can be cast in a similar form:
\begin{equation}
\label{xiosum}
 \tilde\sxi ( \omega_{12} , \tilde\triangle )
  =
 \tilde A_i ( \omega_{12} , \tilde\triangle ) \ff^i \tilde\rxi_{\omega_{12}}
\end{equation}
where $\tilde\triangle$ is shorthand for the `shape' variables
$[ ( \omega_1 \!-\! \omega_2 )/2 , l ]$
at fixed $\omega_{12} = \omega_1 + \omega_2$.
The quantity on the left hand sides
of equations (\ref{xisum}) and (\ref{xiosum})
is the redshift space correlation function, an observable quantity.
On the right hand sides
of equations (\ref{xisum}) and (\ref{xiosum})
is the ensemble average distortion predicted
when the underlying pattern of clustering is
statistically homogeneous and isotropic.
The problem is to find the values of $\ff^i \rxi_{\omega_{12}}$
that give the best fit between observation and prediction.
The ratios of any two of the three fitted quantities
$\rxi_{\omega_{12}}$, $\ff \rxi_{\omega_{12}}$ and $\ff^2 \rxi_{\omega_{12}}$
will then give values for $\ff$ that are independent of $\rxi_{\omega_{12}}$.
It is convenient to think of $\ff$ and $\ff^2$ as `independent'
parameters to be fitted to the data,
which can be combined into a single best fit at the end of the day.

For a sufficiently large survey,
the maximum likelihood solution of equation (\ref{xisum})
at any particular $\omega_{12}$
is the least-squares solution of
(Kendall \& Stuart 1967, section 19.17)
\begin{equation}
\label{chisq}
 \chi^2 =
  \int \!\!\! \int
  [ \sxi ( \omega_{12} , \triangle )
  -  A_i ( \omega_{12} , \triangle ) \ff^i \rxi_{\omega_{12}} ]
  [ \sxi ( \omega_{12} , \triangle ' )
  - A_j ( \omega_{12} , \triangle ' ) \ff^j \rxi_{\omega_{12}} ]
  W ( \omega_{12} , \triangle , \triangle ' )
  {\rmn d} \triangle {\rmn d} \triangle '
\end{equation}
where the weighting function
$W ( \omega_{12} , \triangle , \triangle ' )$
is the inverse of the covariance matrix
\begin{equation}
\label{W}
 W ( \omega_{12} , \triangle , \triangle ' )
  = \langle \Delta \sxi ( \omega_{12} , \triangle )
  \Delta \sxi ( \omega_{12} , \triangle ' ) \rangle ^{-1}
 \ .
\end{equation}
The volume element ${\rmn d} \triangle$ of triangle configurations is
\begin{equation}
 {\rmn d} \triangle = {r_1 {\rmn d} r_1 r_2 {\rmn d} r_2 \over r_{12}^4}
 \ \ \ , \ \ \ \ \ 
\end{equation}
in terms of which the pair-volume element is
${\rmn d}^3 r_1 {\rmn d}^3 r_2
 = 8 \pi^2 {\rmn d} \triangle \, r_{12}^6 {\rmn d} r_{12} / r_{12}$,
the factor $8 \pi^2$ coming from integration over orientations.
Similar equations are valid in the $\omega lm$-representation
(just put tildes on $\sxi$, $A$, $\triangle$ and $W$ in
equations [\ref{chisq}] and [\ref{W}]),
the volume element ${\rmn d} \tilde\triangle$ of configurations being
\begin{equation}
 {\rmn d} \tilde\triangle =
  {\rmn d} ( \omega_1 \!-\! \omega_2 )/2
\end{equation}
with integration implying summation over angular indices $lm$.
For complex $\sxi$, as here, the integrand of equation (\ref{chisq})
should be interpreted as representing its real and imaginary parts
separately, or else some combination thereof,
according to the form of the covariance matrix in equation (\ref{W}).

The sum of squares $\chi^2$, equation (\ref{chisq}),
is minimized when the derivatives
with respect to the parameters $\ff^i \rxi_{\omega_{12}}$ are zero,
$\partial \chi^2 / \partial \ff^i \rxi_{\omega_{12}} = 0$.
This is a set of linear equations, whose solution is
\begin{equation}
\label{xifit}
 \ff^i \rxi_{\omega_{12}} =
  M_{ij}^{-1} ( \omega_{12} )
  \int A_j ( \omega_{12} , \triangle )
  \sxi ( \omega_{12} , \triangle ' )
  W ( \omega_{12} , \triangle , \triangle ' )
  {\rmn d} \triangle {\rmn d} \triangle '
\end{equation}
where $M_{ij}^{-1} ( \omega_{12} )$
is the inverse of the symmetric $3 \times 3$ matrix
\begin{equation}
 M_{ij} ( \omega_{12} ) =
  \int A_j ( \omega_{12} , \triangle )
  A_i ( \omega_{12} , \triangle ' )
  W ( \omega_{12} , \triangle , \triangle ' )
  {\rmn d} \triangle {\rmn d} \triangle '
 \ .
\end{equation}

If the covariance matrix is diagonal,
then it is trivial to invert and hence determine the optimal weighting
function $W$, equation (\ref{W}).
As made clear by
Feldman, Kaiser \& Peacock (1994)
and exploited by Heavens \& Taylor (1995),
this is one of the great advantages of working in Fourier space
for Gaussian fluctuations,
that the independence of (unredshifted) Fourier modes
causes the covariance matrix of the (unredshifted) power to be diagonal.
Feldman et al.'s minimum variance pair weighting
$[ 1 + \Phi P ( k ) ]^{-2}$
is definitive at scales which are Gaussian
but smaller than the scale of the survey.

As argued in Section \ref{waves},
the advantages of mode independence are preserved at least partially
in the $\omega lm$-representation.
We pursue this idea momentarily,
before abandoning it in Section \ref{rrep} in favour of the
$\omega_{12} \triangle$-representation.
The covariance matrix of the correlation function in $\omega lm$-space is
\begin{equation}
\label{xxo}
 \langle \Delta \tilde\sxi_{\omega_1 \omega_2  l m}
  \Delta \tilde\sxi_{\omega_3 \omega_4  l' m'} \rangle
  = \rdelta_{l l'}
  ( \rdelta_{m m'}
    \tilde\sxi_{\omega_1 \omega_4 l}
    \tilde\sxi_{\omega_2 \omega_3 l}
  + \rdelta_{m , - m'}
    \tilde\sxi_{\omega_1 \omega_3 l}
    \tilde\sxi_{\omega_2 \omega_4 l} )
  + \tilde\seta_{\omega_1 \omega_2 \omega_3 \omega_4  l m l' m'}
\end{equation}
where $\tilde\seta$ is the four-point function,
which vanishes for Gaussian fluctuations,
and is neglected hereafter.
For the unredshifted correlation function, equation (\ref{xikred}),
the covariance matrix
(\ref{xxo}) reduces to
\begin{equation}  
\label{xxor}
 \langle \Delta \tilde\rxi_{\omega_1 \omega_2  l m}
  \Delta \tilde\rxi_{\omega_3 \omega_4  l' m'} \rangle
  = \rdelta_{l l'}
  ( \rdelta_{m m'}
    \tilde\rxi_{\omega_{14}}
    \tilde\rxi_{\omega_{23}}
  + \rdelta_{m , - m'}
    \tilde\rxi_{\omega_{13}}
    \tilde\rxi_{\omega_{24}} )
 \ .
\end{equation}
For a power-law power spectrum,
the unredshifted correlation function
in $\omega lm$-space is a Dirac delta-function,
$\tilde\rxi_{\omega_{12}} = \delta_{\rmn D} ( \omega_{12} )$.
In this case,
the only non-zero elements of the covariance matrix
(\ref{xxor})
are the diagonal elements
$\langle \Delta \tilde\rxi_{\omega_1 \omega_2  l m}
\Delta \tilde\rxi^{\ast}_{\omega_1 \omega_2  l m} \rangle$,
and these elements are all equal.
In effect, the covariance matrix is the identity matrix.
The covariance matrix of the redshifted correlation function
is related to its unredshifted counterpart
by the spherical distortion equation (\ref{bxiol}).
In redshift space,
the covariance matrix (\ref{xxo})
remains diagonal for a power-law power spectrum,
but it depends on $\ff$.

The above argument,
that the covariance matrix (\ref{xxo}) is almost diagonal in
the $\omega lm$-representation,
contains a serious flaw.
Namely, it neglects to take into account the shot noise
caused by the discrete sampling of galaxies.
Shot noise destroys scale invariance,
and introduces off-diagonal correlations between modes.
We are not sure whether this difficulty is fatal.
Whatever the case, it is enough to persuade us
to switch to the $\omega_{12} \triangle$-representation,
which proves more tractable.

\subsection{$\omega_{12} \triangle$-representation}
\label{rrep}

Rather than attempt a weighting that is absolutely optimal,
let us instead adopt a weighting function $W$
that is diagonal in real space.
Great precision in the choice of $W$ is not required,
since any weighting in the vicinity of minimum variance should
give a variance not much different from the minimum.
Thus we seek to minimize at each separation $r_{12}$ the sum of squares
\begin{equation}
\label{chisqr}
 \chi^2 =
  \int
  \left[ \sxi ( r_{12} , \triangle )
  - \int_{-\infty}^{\infty}
  A_i ( \omega_{12} , \triangle ) \ff^i \rxi_{\omega_{12}}
  r_{12}^{- \gamma - i \omega_{12}} {\rmn d} \omega_{12}
  \right]^2
  W ( r_{12} , \triangle )
  {\rmn d} \triangle
 \ .
\end{equation}
A near-minimum variance form of
the weighting function $W ( r_{12} , \triangle )$
is (Hamilton 1993b, section 5;
notice the following formula is a volume weighting
not a number weighting, which accounts for the $\Phi ( r_1 ) \Phi ( r_2 )$
in the numerator)
\begin{equation}
 W ( r_{12} , \triangle ) =
  {\Phi ( r_1 ) \Phi ( r_2 )
  \over [ 1 + \Phi ( r_1 ) J ( r_{12} ) ] [ 1 + \Phi ( r_2 ) J ( r_{12} ) ]}
\end{equation}
where $J ( r_{12} ) \approx \int_0^{r_{12}} \xi {\rmn d} V$.

According to
equations (\ref{A1}) and (\ref{A2}),
the coefficients $A_i ( \omega_{12} , \triangle )$
separate into sums of products
of the five shape-functions $B_j ( \triangle )$,
equations (\ref{B0})--(\ref{B4}),
with functions $a_{ji} ( \omega_{12} )$
which depend on $\omega_{12}$ but not on triangle shape $\triangle$:
\begin{equation}
 A_i ( \omega_{12} , \triangle ) = B_j ( \triangle ) a_{ji} ( \omega_{12} )
 \ .
\end{equation}
It is convenient to imagine
the five quantities
$\int_{-\infty}^{\infty}
  a_{ji} ( \omega_{12} ) \ff^i \rxi_{\omega_{12}}
  r_{12}^{- \gamma - i \omega_{12}} {\rmn d} \omega_{12}$
as `independent' parameters to be fitted to the data.
Minimizing the sum of squares (\ref{chisqr})
with respect to these parameters yields five equations
(cf.\ equation [\ref{xifit}]):
\begin{equation}
\label{xifitr}
 M_{ij}^{-1} ( r_{12} )
 \int B_i ( \triangle ) \sxi ( r_{12} , \triangle ) W ( r_{12} , \triangle )
  {\rmn d} \triangle =
  \int_{-\infty}^{\infty}
  a_{ji} ( \omega_{12} ) \ff^i \rxi_{\omega_{12}}
  r_{12}^{- \gamma - i \omega_{12}} {\rmn d} \omega_{12}
\end{equation}
where $M_{ij}^{-1} ( r_{12} )$ is the inverse of the symmetric
$5 \times 5$ matrix
\begin{equation}
\label{Mij}
 M_{ij} ( r_{12} ) =
 \int B_i ( \triangle ) B_j ( \triangle ) W ( r_{12} , \triangle ) 
  {\rmn d} \triangle
 \ .
\end{equation}
Define $\sXi_i ( r_{12} )$
to be the vector on the left hand side of equation (\ref{xifitr}):
\begin{equation}
\label{Xi}
 \sXi_i ( r_{12} ) \equiv
  M_{ij}^{-1} ( r_{12} )
  \int B_j ( \triangle ) \sxi ( r_{12} , \triangle ) W ( r_{12} , \triangle ) 
  {\rmn d} \triangle
 \ .
\end{equation}
As will become apparent below,
the $\sXi_i ( r_{12} )$ can be interpreted as the generalization
to spherical distortions of the harmonics of the redshift correlation function
in the plane-parallel case.
On the right hand side of equation (\ref{xifitr}),
the quantities $a_{ji} ( \omega_{12} )$ can be replaced by an
operator in real space and taken outside the integral.
Equation (\ref{xifitr}) then becomes, explicitly,
\begin{equation}
\label{Xi0}
 \sXi_0 ( r_{12} ) = \left(
  1 + \frac{2}{3} \ff + \frac{1}{5} \ff^2
  \right) \rxi ( r_{12} )
\end{equation}
\begin{equation}
\label{Xi2}
 \sXi_2 ( r_{12} ) = \left(
  \frac{4}{3} \ff + \frac{4}{7} \ff^2
  \right) \left[ \rxi ( r_{12} ) - \rxibar ( r_{12} ) \right]
\end{equation}
\begin{equation}
\label{Xi4}
 \sXi_4 ( r_{12} ) =
  \frac{8}{35} \ff^2
  \left[ \rxi ( r_{12} ) + \frac{5}{2} \rxibar ( r_{12} )
  - \frac{7}{2} \rxibarbar ( r_{12} ) \right]
\end{equation}
\begin{equation}
\label{Xi1}
 \sXi_1 ( r_{12} ) =
  \frac{1}{3} \ff \alpha \rxibar ( r_{12} )
  + \ff^2 \left\{
   \frac{1}{5} \alpha \rxibarbar ( r_{12} )
   + \alpha^2 \left[ \frac{1}{6} \breve\rxi ( r_{12} )
                - \frac{1}{15} \rxibarbar ( r_{12} ) \right]
  \right\}
\end{equation}
\begin{equation}
\label{Xi3}
 \sXi_3 ( r_{12} ) =
  \ff^2 \left\{
   \frac{2}{3} \rxibar ( r_{12} ) - \frac{4}{5} \rxibarbar ( r_{12} )
   - \alpha \left[ \frac{1}{6} \rxibar ( r_{12} )
            + \frac{1}{10} \rxibarbar ( r_{12} ) \right]
   - \alpha^2 \left[ \frac{1}{3} \breve\rxi ( r_{12} )
                - \frac{1}{6} \rxibar ( r_{12} )
                - \frac{1}{30} \rxibarbar ( r_{12} ) \right]
  \right\}
\end{equation}
where, following Hamilton's (1992) notation,
\begin{equation}
 \breve\rxi ( r_{12} ) \equiv
  2 r_{12}^{-2} \int_0^{r_{12}} \rxi ( r ) r {\rmn d} r
\ \ \ , \ \ \ \ \ 
 \rxibar ( r_{12} ) \equiv
  3 r_{12}^{-3} \int_0^{r_{12}} \rxi ( r ) r^2 {\rmn d} r
\ \ \ , \ \ \ \ \ 
 \rxibarbar ( r_{12} ) \equiv
  5 r_{12}^{-5} \int_0^{r_{12}} \rxi ( r ) r^4 {\rmn d} r
\ \ \ .
\end{equation}

In the plane-parallel limit,
$\sXi_0$, $\sXi_2$ and $\sXi_4$ go over to the monopole,
quadrupole and hexadecapole harmonics of the correlation function,
while $\sXi_1$ and $\sXi_3$ vanish.
The three equations (\ref{Xi0})--(\ref{Xi4}) for
$\sXi_0$, $\sXi_2$ and $\sXi_4$
look exactly like their plane-parallel counterparts,
equations (6)--(8) of Hamilton (1992).
In particular, the `quadrupole' to `monopole' ratio
\begin{equation}
\label{Xi20}
 {\sXi_2 ( r_{12} ) \over \sXi_0 ( r_{12} ) - \overline\sXi_0 ( r_{12} )}
  = {{4 \over 3} \ff + {4 \over 7} \ff^2
  \over 1 + {2 \over 3} \ff + {1 \over 5} \ff^2}
\end{equation}
provides a way to measure $\ff$ in a manner independent of the
shape of the correlation function,
but now for fully spherical distortions.

As remarked by Cole et al.\ (1994, appendix B),
the combinations of $\rxi$ on the right hand sides of equations
(\ref{Xi0})--(\ref{Xi4})
can be regarded as arising from windowing the power spectrum
with spherical Bessel functions $j_l ( k r_{12} )$.
Thus if $\tildesXi_l ( k )$ for $l = 0$, 2, 4 are defined by
\begin{equation}
\label{P}
 \tildesXi_l ( k ) =
  4 \pi {\rmn i}^l
  \int_0^{\infty} \sXi_l ( r_{12} ) j_l ( k r_{12} ) r_{12}^2 {\rmn d} r_{12}
\ \ \ , \ \ \ \ \ 
 \sXi_l ( r_{12} ) =
  {1 \over 2 \pi^2 {\rmn i}^l}
  \int_0^{\infty} \tildesXi_l ( k ) j_l ( k r_{12} ) k^2 {\rmn d} k
\end{equation}
then the three equations (\ref{Xi0})--(\ref{Xi4}) become
\begin{equation}
\label{P0}
 \tildesXi_0 ( k ) = \left(
  1 + \frac{2}{3} \ff + \frac{1}{5} \ff^2
  \right) P ( k )
\end{equation}
\begin{equation}
\label{P2}
 \tildesXi_2 ( k ) = \left(
  \frac{4}{3} \ff + \frac{4}{7} \ff^2
  \right) P ( k )
\end{equation}
\begin{equation}
\label{P4}
 \tildesXi_4 ( k ) =
  \frac{8}{35} \ff^2
  P ( k )
\end{equation}
where $P ( k )$ is the unredshifted power spectrum.
Hence the `quadrupole' to `monopole' ratio (\ref{Xi20}) can also be written
\begin{equation}
\label{P20}
 {\tildesXi_2 ( k ) \over \tildesXi_0 ( k )}
  = {{4 \over 3} \ff + {4 \over 7} \ff^2
  \over 1 + {2 \over 3} \ff + {1 \over 5} \ff^2}
\ .
\end{equation}

The presence of the inverse $5 \times 5$ matrix $M_{ij}^{-1} ( r_{12} )$
in equation (\ref{Xi})
makes the quantities $\sXi_i ( r_{12} )$
more difficult to evaluate than in the plane-parallel limit.
However,
the $5 \times 5$ elements of the matrix, equation ({\ref{Mij}),
are integrals over known quantities
$B_i ( \triangle )$
and the chosen weighting function $W ( r_{12} , \triangle )$,
so can be precomputed.

\subsection{The validity of the constant $\alpha$ approximation}
\label{nalph}

Equations (\ref{Xi0})--(\ref{Xi3}) are valid provided
that $\alpha ( r )$, equation (\ref{alpha}), is approximated
as a constant (cf.\ Section \ref{alph}).
How can the validity of this approximation be checked?

Mathematically,
the passage from equation (\ref{chisqr}) via equation (\ref{Xi})
to equations (\ref{Xi0})--(\ref{Xi3})
involves the approximation that
\begin{equation}
\label{aM}
 \int \alpha ( r_1 )
  B_i ( \triangle ) B_j ( \triangle ) W ( r_{12} , \triangle ) 
  {\rmn d} \triangle
  = \alpha M_{ij} ( r_{12} )
\end{equation}
where
$\alpha$ on the right hand side is required to be a constant independent
of $r_{12}$ and $ij$.
A similar constraint on $\alpha ( r_1 ) \alpha ( r_2 )$ is required.
Note that $\alpha ( r )$ in equations (\ref{A1}) and (\ref{A2})
multiplies only the shape functions $B_1$ and $B_3$,
so it suffices that equation (\ref{aM}) be valid when either of
$i$ or $j$ is 1 or 3.
The validity of equations (\ref{aM}) should be checked
in applying equations (\ref{Xi0})--(\ref{Xi3}).

\subsection{Deconvolving the catalogue window}

In measuring the redshift space correlation function
$\sxi ( r_{12} , \triangle )$,
it is essential to disentangle the true anisotropy of the
redshift correlation function
from the artificial anisotropy induced by the angular and radial
selection functions of the survey.
This deconvolution is easy in real space,
where the observed galaxy density is the product of the true density
and the selection function of the survey,
but is complicated in any other space.

The `deconvolution' procedure is described in some detail by
Hamilton (1993b, section 6)
for the plane-parallel case where the aim is to
measure the redshift correlation function
$\sxi ( r_{12} , \mu )$
as a function of separation $r_{12}$ and cosine angle $\mu$ to the
line of sight.
The procedure generalizes easily to the present case
where the redshift correlation function
$\sxi ( r_{12} , \triangle )$
is a function of separation $r_{12}$ and triangle shape $\triangle$.
Briefly,
each observed galaxy pair
at separation $r_{12}$ and triangle configuration $\triangle$
makes a contribution to the correlation function
$\sxi ( r_{12} , \triangle )$.
To correct for the selection function of the survey,
one simply divides by the probability of finding a pair
at $r_{12}$ and $\triangle$,
given the boundaries and selection function of the survey.
\section{Conclusions}

The aim of this paper has been to
present the theoretical foundation of a procedure for measuring
the linear growth rate $\ff$,
hence the cosmological density $\Omega$ in unbiased standard cosmology,
from spherical redshift space distortions
in a manner independent of the shape of the power spectrum.

Our most important point, presented in Section 3,
is that there exists an operator $\ddlnr$
which both commutes with the spherical distortion operator,
and defines a characteristic scale of separation $r_{12}$.
Ratios of the amplitudes of the eigenfunctions of this operator
provide measures of $\ff$ independent of the shape of the power spectrum.

In Section 4,
we presented a complete set of commuting operators for the spherical
distortion operator.
The eigenfunctions of this complete set are spherical waves about the observer,
with radial part lying in logarithmic real or Fourier space.

In Section 5,
we discussed the practical measurement of $\ff$
using the ideas presented.
In particular,
we showed that there is a set of five functions $\sXi_i ( r_{12} )$,
equations (\ref{Xi0})--(\ref{Xi3}),
which can be regarded as the generalization to fully spherical distortions
of the monopole, quadrupole, and hexadecapole harmonics
of the correlation function in the plane-parallel case.

A drawback of the method is that
the spherical distortion operator commutes with $\ddlnr$,
and likewise with the complete set of operators in Section 4,
only to the extent that the logarithmic slope of the radial
selection function can be approximated by a constant.
We argued in Section \ref{alph} that this may be a reasonable approximation
in practice,
and we showed in Section \ref{nalph} how to check the validity
of the approximation.

\section*{Acknowledgments}

This work was supported by
NSF grant AST93-19977,
NASA Astrophysical Theory Grant NAG 5-2797,
and a PPARC Visiting Fellowship (AJSH).
We thank George Efstathiou for the hospitality of the
Nuclear and Astrophysics Laboratory at Oxford University,
where much of this work was done.

\bsp
\end{document}